\documentclass[a4paper,fleqn,usenatbib]{mnras}
\usepackage[T1]{fontenc}
\usepackage{ae,aecompl}

\usepackage{graphicx}	% Including figure files
\usepackage{amsmath}	% Advanced maths commands
\usepackage{amssymb}	% Extra maths symbols

\usepackage{url}

\title[Testing difference-smoothing on TDC1 simulations]{Crash testing difference-smoothing algorithm on a large sample of simulated light curves from TDC1}

\author[S. Rathna Kumar]{
S. Rathna Kumar$^{1,2}$\thanks{E-mail: rathna@prl.res.in}
\\
$^{1}$Physical Research Laboratory, Navrangpura, Ahmedabad 380009, India\\
$^{2}$Aryabhatta Research Institute of Observational Sciences, Manora Peak, Nainital 263002, India 
}

\date{Accepted 2017 May 31. Received 2017 May 28; in original form 2017 March 27}

\pubyear{2017}

\begin{document}
\label{firstpage}
\pagerange{\pageref{firstpage}--\pageref{lastpage}}
\maketitle

\begin{abstract}
In this work, we propose refinements to the difference-smoothing algorithm for measurement of time delay from the light curves of the images of a gravitationally lensed quasar. The refinements mainly consist of a more pragmatic approach to choose the smoothing time-scale free parameter, generation of more realistic synthetic light curves for estimation of time delay uncertainty and using a plot of normalized $\chi^2$ computed over a wide range of trial time delay values to assess the reliability of a measured time delay and also for identifying instances of catastrophic failure. We rigorously tested the difference-smoothing algorithm on a large sample of more than thousand pairs of simulated light curves having known true time delays between them from the two most difficult `rungs' -- rung3 and rung4 -- of the first edition of Strong Lens Time Delay Challenge (TDC1) and found an inherent tendency of the algorithm to measure the magnitude of time delay to be higher than the true value of time delay. However, we find that this systematic bias is eliminated by applying a correction to each measured time delay according to the magnitude and sign of the systematic error inferred by applying the time delay estimator on synthetic light curves simulating the measured time delay. Following these refinements, the TDC performance metrics for the difference-smoothing algorithm are found to be competitive with those of the best performing submissions of TDC1 for both the tested `rungs'. The MATLAB codes used in this work and the detailed results are made publicly available.   
\end{abstract}


%%%%%%%%%%%%%%%%%%%%%%%%%%%%%%%%%%%%%%%%%%%%%%%%%%

%%%%%%%%%%%%%%%%% BODY OF PAPER %%%%%%%%%%%%%%%%%%

\section{Introduction}

Strong gravitational lensing occurs when a sufficiently massive galaxy or a galaxy cluster lies in close proximity to the line of sight of a distant background source, leading to the observer seeing multiple distorted images of the background source. The individual images are magnified in flux relative to one another and also with respect to the actual image of the background source, which cannot be seen. Similarly, the images are delayed in arrival time with respect to one another due to geometric differences in the light travel paths and also due to the paths traversing different regions of the gravitational potential of the massive deflector \citep[e.g.][]{Treu2010}. When the background source is variable in flux such as quasar or supernova explosion, it is possible to measure the time delays between the individual images by monitoring their brightness variations and matching the variability features in their light curves \citep[e.g.][]{Tewes2013,Rodney2016}. These time delays in combination with modelling of the mass distribution of the deflector can be used to constrain cosmological parameters, mainly $H_0$ \citep[e.g.][]{Suyu2010,Suyu2013,Bonvin2017}. The idea was originally proposed by \citet{Refsdal1964} even before the discovery of the first gravitational lens \citep*{Walsh1979}.

Measurement of time delays between the images of a gravitationally lensed quasar is non-trivial due to the irregular sampling of the light curves arising from telescope scheduling and weather constraints and the presence of large gaps during non-visibility periods of the object \citep*[e.g.][]{Hojjati2013,Tewes2013a}. A further complication is the possible presence of extrinsic variations in the light curves due to microlensing by stars in the lensing galaxy, which are uncorrelated between the light curves of the different images \citep{Chang1979}. Whereas currently time delays have been reported in approximately two dozen lensed quasars \citep*[e.g.][]{Kumar2015}, during the next decade with the advent of Large Synoptic Survey Telescope\footnote{\url{https://www.lsst.org/}} (LSST), the number of systems with monitoring light curves spanning several years is expected to be $\sim$10$^3$ \citep[see][and references therein]{Treu2016}. Hence it is of interest to develop time delay measurement techniques that are fast, yet at the same time accurate and precise \citep[e.g.][]{Hojjati2014,Bonvin2016}. To assess the present day capabilities of the community as far as measurement of time delays from lensed quasar light curves with LSST-like sampling properties are concerned and also to provide inputs for finalizing the exact survey strategy that will be adopted by LSST, a team of scientists from the Dark Energy Science Collaboration invited the community members to participate in a Strong Lens Time Delay Challenge\footnote{\url{http://timedelaychallenge.org/}} \citep[TDC;][]{Dobler2015}, which consisted of two `ladders', TDC0 and TDC1. One of the seven teams which qualified for TDC1 employed the difference-smoothing technique \citep{Liao2015}.

However, our TDC1 submission based on difference-smoothing was unsatisfactory for several reasons. Only a small fraction of the TDC1 light curves could be analysed due to poor automation of our codes implementing the algorithm. Despite following an extensive procedure to estimate the uncertainties of the measured time delays, our submission still suffered from the presence of catastrophic outliers. The accuracy and precision metrics were poor not only due to the smallness of the analysed sample but also due to not performing any selection to separate the high quality measurements from the low quality ones. In this work, we focus on introducing refinements to the difference-smoothing technique for measurement of time delay from the light curves of lensed quasar images. As a result of improved automation of our codes, we are now able to test the refined procedure on a large sample of more than thousand pairs of simulated light curves with known true time delays between them from the two most difficult `rungs' -- rung3 and rung4 -- of TDC1. This paper is organized as follows. In Section~\ref{sec:difference-smoothing}, we recall the difference-smoothing technique and propose the refinements to the procedure followed for measurement of time delay and estimation of its uncertainty, and explain the motivations for these changes. In Section~\ref{sec:crash-testing}, we describe the TDC performance metrics and then present the results of rigorously testing the revised procedure on a large sample of TDC1 simulated light curves. We briefly conclude in Section~\ref{sec:conclusion}.   

\section{Difference-smoothing technique}
\label{sec:difference-smoothing}

The difference-smoothing technique as introduced in \citet{Kumar2013}, in the context of measurement of time delay of the doubly lensed quasar SDSS J1001+5027, is a point estimator that determines both an optimal time delay and an optimal flux shift between two light curves, while at the same time allowing for smooth extrinsic variability. It is based on comparing the difference light curve with a smoothed version of it and minimizing the residuals. In \citet{Kumar2015}, we modified the technique such that it no longer performed a flux shift between the light curves. This modification was  incorporated to make it computationally less time-consuming. The basic technique for measurement of time delay followed in this work remains the same as in \citet{Kumar2015}. However we briefly review the technique here not only for the convenience of the reader but also to introduce the notation needed to describe the proposed refinements to the procedure followed for measurement of time delay and for estimation of uncertainty of the measured time delay. 

\subsection{Measurement of time delay}
\label{sec:time-delay}

We have light curves $A$ and $B$ consisting of magnitudes $A_i$ and $B_i$, respectively, at observing epochs $t_i$ $(i=1,2,3,...,N)$, arranged in increasing order of time. The magnitudes $A_i$ and $B_i$ have photometric errors $\sigma_{A_i}$ and $\sigma_{B_i}$, respectively. We form the difference light curve $d_i$ for a trial time delay $\tau$ as 
\begin{equation}
    d_i(\tau)=A_i-\frac{\sum_{j=1}^N w_{ij}B_j^\prime}{\sum_{j=1}^N w_{ij}},
	\label{eq:diff-lightcurve}
\end{equation}
where $B_j^\prime$ represent the magnitudes of $B^\prime$, the time shifted version of the $B$ light curve, having observing epochs $t_j^\prime=t_j-\tau$. In the present context where we do not perform any flux shift, $B_j^\prime = B_j$. The weights $w_{ij}$ are given by
\begin{equation}
    w_{ij}=\frac{\mathrm{e}^{-(t_j^\prime-t_i)^2/2\delta^2}}{\sigma_{B_j}^2},
	\label{eq:weights-w}
\end{equation}
where $\delta$ called as decorrelation length is one of the free parameters of the technique. The uncertainty of $d_i$ is computed as
\begin{equation}
    \sigma_{d_i}=\sqrt{\sigma_{A_i}^2+\frac{1}{\sum_{j=1}^N w_{ij}}}.
	\label{eq:sigma-d_i}
\end{equation} 
A smoothed version of the difference light curve $d_i$ is obtained as
\begin{equation}
    f_i(\tau)=\frac{\sum_{j=1}^N \nu_{ij}d_j}{\sum_{j=1}^N \nu_{ij}}.
	\label{eq:f_i}
\end{equation} 
The weights $\nu_{ij}$ are given by
\begin{equation}
    \nu_{ij}=\frac{\mathrm{e}^{-(t_j-t_i)^2/2s^2}}{\sigma_{d_j}^2},
	\label{eq:weights-nu}
\end{equation}
where the smoothing time-scale $s$ is a second free parameter of the technique. $f_i$ represents a model of the differential extrinsic variation for the trial time delay $\tau$. The uncertainty of $f_i$ is computed as
\begin{equation}
    \sigma_{f_i}=\sqrt{\frac{1}{\sum_{j=1}^N \nu_{ij}}}.
	\label{eq:sigma-f_i}
\end{equation} 
The time delay $\Delta t$ is found by optimizing the trial time delay $\tau$ to minimize the residuals between $d_i$ and its smoothed version $f_i$. We achieve this by minimizing a normalized $\chi^2$ defined as
\begin{equation}
    \overline{\chi}^2(\tau)=\left[\sum_{i=1}^N \frac{(d_i-f_i)^2}{\sigma_{d_i}^2+\sigma_{f_i}^2}\right]/\left[\sum_{i=1}^N \frac{1}{\sigma_{d_i}^2+\sigma_{f_i}^2}\right].
	\label{eq:chi-squared}
\end{equation}
Since the above process uses $A$ light curve as reference, we repeat the calculation of $\overline{\chi}^2$ for each trial time delay using $B$ light curve as reference. We average the two values of $\overline{\chi}^2$ and minimize this average value to find the time delay $\Delta t$. We note that in the present work, we adopt the TDC convention of $\Delta t > 0$ corresponding to light curve $A$ leading light curve $B$. 

\subsection{Generation of synthetic light curves}
\label{sec:synthetic-lightcurves} 

In \citet{Kumar2013}, to find the uncertainty of the measured time delay, we followed the Monte Carlo analysis described in \citet{Tewes2013a}, which consists of applying the point estimator to a large number of realistic synthetic light curves, which closely mimic the properties of the observed data, covering a range of simulated time delays around a plausible solution. In \citet{Kumar2015}, we introduced an independent recipe for generating synthetic light curves having the same properties as the observed light curves with simulated time delays at discrete values in a plausible range around the measured time delay. We also introduced a reasonable scheme for setting the values of the two free parameters of the difference-smoothing technique according to the properties of the light curves from which we are measuring the time delay. We then tested the entire procedure on a sample of 250 publicly available pairs of simulated light curves from TDC1 with known true time delay values, fifty from each of the five `rungs', selected such that they were of sufficiently good quality for measurement of time delay. For all except one pair among those 250 TDC1 simulated light curves, the measured time delays agreed with the true time delays to within 2$\sigma_i$. The exceptional case had a measured time delay that was discrepant with the true time delay at the level of 2.25$\sigma_i$. As a result of recent improvements in automation of our codes, we could apply the time delay measurement and uncertainty estimation procedure of \citet{Kumar2015} on a much larger sample of TDC1 simulated light curves, and we encountered many more cases wherein the measured time delays were in tension with the true time delays at >2$\sigma_i$ level. However for the time delay measurement and uncertainty estimation procedure to be considered robust, it is reasonable to expect that all the measured time delays need to match with the true time delays to within $\sim$2$\sigma_i$. 

Based on the above consideration, we introduce refinements to the procedure for generating synthetic light curves that are used for estimating the uncertainty of the measured time delay. These refinements are aimed at making the synthetic light curves a more realistic representation of the actual light curves from which the time delay is measured. We first identify the individual observing seasons in the light curves $A$ and $B$ by finding those spacings between adjacent observing epochs which are larger than a certain threshold. For each observing epoch $t_i$, we estimate the local mean sampling $m(t_i)$ by averaging $n_s$ spacings between adjacent observing epochs $t_{k+1}-t_k$ within the same observing season, whose centres $(t_k+t_{k+1})/2$ are nearest to $t_i$. We now proceed to obtain empirical estimates of noise in the light curves $A$ and $B$, based on the local scatter properties of the light curves \citep{Tewes2013a}. For all epochs $t_i$, we calculate the residuals of the magnitudes with respect to a model of the underlying variation inferred based on the magnitudes of all the epochs as   
\begin{equation}
    r_A(t_i)=A_i-A(t_i)
	\label{eq:residuals-A}
\end{equation}
and  
\begin{equation}
    r_B(t_i)=B_i-B(t_i),
	\label{eq:residuals-B}
\end{equation}
where
\begin{equation}
    A(t_i)=\frac{\sum_{j=1}^N \mathrm{e}^{-(t_j-t_i)^2/2m(t_i)^2} A_j}{\sum_{j=1}^N \mathrm{e}^{-(t_j-t_i)^2/2m(t_i)^2}}
	\label{eq:smoothed-A}
\end{equation}
and  
\begin{equation}
     B(t_i)=\frac{\sum_{j=1}^N \mathrm{e}^{-(t_j-t_i)^2/2m(t_i)^2} B_j}{\sum_{j=1}^N \mathrm{e}^{-(t_j-t_i)^2/2m(t_i)^2}},
	\label{eq:smoothed-B}
\end{equation}
for the $A$ and $B$ light curves, respectively. Now for each epoch $t_i$, to obtain empirical estimates of noise $\hat{\sigma}_{A_i}$ and $\hat{\sigma}_{B_i}$, we take the standard deviation of $n_r$ number of $r_A$ and $r_B$ residuals, respectively, within the same observing season whose epochs $t_k$ are nearest to $t_i$. In this work, we set $n_s$ = 10 and $n_r$ = 10, these values being chosen to be large enough for $m(t_i)$, $\hat{\sigma}_{A_i}$ and $\hat{\sigma}_{B_i}$ to be well behaved, while at the same time being small enough for the synthetic light curves to adequately mimic the local properties of the actual observed light curves (see Fig.~\ref{fig:light-curves}).  

We note the differences in the above procedure to obtain $\hat{\sigma}_{A_i}$ and $\hat{\sigma}_{B_i}$ from that followed in \citet{Kumar2015}, where we had performed a uniform rescaling of the stated photometric errors $\sigma_{A_i}$ and $\sigma_{B_i}$. We no longer use factors dependent on the photometric errors in assigning weights for the different terms in equation~(\ref{eq:smoothed-A}) and equation~(\ref{eq:smoothed-B}), thus making the empirical estimates of noise $\hat{\sigma}_{A_i}$ and $\hat{\sigma}_{B_i}$ completely independent of $\sigma_{A_i}$ and $\sigma_{B_i}$. Also, instead of using a single value of mean sampling $m$ estimated for the entire light curve after excluding large seasonal gaps, we now use a value of mean sampling $m(t_i)$ estimated locally for each observing epoch. 

We merge the two light curves by shifting the $B$ light curve by the measured time delay $\Delta t$ and subtracting the model of differential extrinsic variations $f_i$ from the $A$ light curve. The merged light curve $M_i$ consists of $A_i-f_i$ and $B_i$ at epochs $t_i$ and  $t_i-\Delta t$, respectively, and from this merged light curve, a model of the quasar brightness variation $M(t)$ is inferred as
\begin{equation}
    M(t)=\frac{\sum_{j=1}^{2N} \frac{1}{\sigma_{M_j}^2}\mathrm{e}^{-(t_j-t)^2/2m(t)^2} M_j}{\sum_{j=1}^{2N} \frac{1}{\sigma_{M_j}^2}\mathrm{e}^{-(t_j-t)^2/2m(t)^2}},
	\label{eq:brightness-variation}
\end{equation}
where $m(t) = m(t_k)$ with the value of $t_k$ being chosen to be that epoch of $M_i$ which is nearest to $t$, and $\sigma_{M_j}$ consists of $\hat{\sigma}_{A_j}$ and $\hat{\sigma}_{B_j}$ at epochs $t_j$ and $t_j-\Delta t$, respectively. Similarly, we model the quasar brightness variation at epochs $t_i$ using only the $A$ points in $M_i$ as
\begin{equation}
    M_A(t_i)=\frac{\sum_{j=1}^N \frac{1}{\hat{\sigma}_{A_j}^2}\mathrm{e}^{-(t_j-t_i)^2/2m(t_i)^2}(A_j-f_j)}{\sum_{j=1}^N \frac{1}{\hat{\sigma}_{A_j}^2}\mathrm{e}^{-(t_j-t_i)^2/2m(t_i)^2}}
	\label{eq:brightness-variation-A}
\end{equation} 
and at epochs $t_i-\Delta t$ using only the $B$ points in $M_i$ as 
\begin{equation}
    M_B(t_i-\Delta t)=\frac{\sum_{j=1}^N \frac{1}{\hat{\sigma}_{B_j}^2}\mathrm{e}^{-(t_j-t_i+\Delta t)^2/2m(t_i)^2}B_j}{\sum_{j=1}^N \frac{1}{\hat{\sigma}_{B_j}^2}\mathrm{e}^{-(t_j-t_i+\Delta t)^2/2m(t_i)^2}}.
	\label{eq:brightness-variation-B}
\end{equation} 
The residual extrinsic variations present in light curves $A$ and $B$, to be incorporated in the synthetic light curves, are calculated as
\begin{equation}
    f_{A_i}=M_A(t_i)-M(t_i)
	\label{eq:residual-extrinsic-variation-A}
\end{equation}    
and
\begin{equation}
    f_{B_i}=M_B(t_i-\Delta t)-M(t_i-\Delta t),
	\label{eq:residual-extrinsic-variation-B}
\end{equation}
respectively. 

In order to generate synthetic light curves $A_i^{\mathrm{simu}}$ and $B_i^{\mathrm{simu}}$, simulating a time delay of $\Delta t+\delta(\Delta t)$ between them, we sample the model of quasar brightness variation $M(t)$ at appropriate values of $t$ and add terms for extrinsic variations and observational noise as follows:
\begin{equation}
    A_i^{\mathrm{simu}}=M\left(t_i+\frac{\delta(\Delta t)}{2}\right)+f_i+f_{A_i}+N^*(0,1)\hat{\sigma}_{A_i}
	\label{eq:simulated-A}
\end{equation} 
and 
\begin{equation}
    B_i^{\mathrm{simu}}=M\left(t_i-\Delta t-\frac{\delta(\Delta t)}{2}\right)+f_{B_i}+N^*(0,1)\hat{\sigma}_{B_i},
	\label{eq:simulated-B}
\end{equation} 
where $N^*(0,1)$ denotes a Gaussian distributed random variate having mean 0 and standard deviation 1. The inclusion of the terms $f_{A_i}$ and $f_{B_i}$ ensure that the synthetic light curves contain short time-scale extrinsic variations \citep{Tewes2013a} in addition to long time-scale extrinsic variations, which is represented by the term $f_i$. The simulated light curves $A_i^{\mathrm{simu}}$ and $B_i^{\mathrm{simu}}$ are both assigned the observing epochs $t_i$ and the photometric errors $\sigma_{A_i}$ and $\sigma_{B_i}$, respectively, as the original light curves. Since the above description for generating synthetic light curves uses $A$ light curve as reference, we repeat the procedure using $B$ light curve as reference and average the corresponding values of $A_i^{\mathrm{simu}}$ and $B_i^{\mathrm{simu}}$ before the addition of the noise terms $N^*(0,1)\hat{\sigma}_{A_i}$ and $N^*(0,1)\hat{\sigma}_{B_i}$ in equation~(\ref{eq:simulated-A}) and equation~(\ref{eq:simulated-B}), respectively.

\subsection{Choosing the values of free parameters}
\label{sec:free-parameters}

The difference-smoothing technique has two free parameters -- the decorrelation length $\delta$ and the smoothing time-scale $s$. The value of $\delta$ is set equal to the mean sampling of the light curves computed after excluding the seasonal gaps, which we shall denote as $m$. The value of $s$ needs to be chosen to be significantly larger than $\delta$. In \citet{Kumar2015}, we had optimized its value so that the maximum of $\left|\frac{f_{A_i}}{\hat{\sigma}_{A_i}}\right|$ and $\left|\frac{f_{B_i}}{\hat{\sigma}_{B_i}}\right|$ was equal to two. These absolute ratios quantify the residual extrinsic variations in units of empirically estimated noise. Here again the maximum absolute ratios for the $A$ and $B$ light curves are computed first with light curve $A$ as reference and then with light curve $B$ as reference and the corresponding values are averaged. The above scheme to optimize the smoothing time-scale free parameter $s$ ensures that its value is set low enough to adequately model the differential extrinsic variations so that the remaining residual extrinsic variations are not significantly larger than the empirically estimated noise present in the light curves. In the absence of there being significant differential extrinsic variations in the light curves, in which case the maximum of $\left|\frac{f_{A_i}}{\hat{\sigma}_{A_i}}\right|$ and $\left|\frac{f_{B_i}}{\hat{\sigma}_{B_i}}\right|$ would be <2 even for large values of $s$, we had set $s=\infty$.

In this work, we propose a more pragmatic approach to choose the value of the smoothing time-scale free parameter $s$. In testing the difference-smoothing algorithm on a large sample of TDC1 simulated light curves, we encountered cases where, even in the absence of a significant amount of extrinsic variations in the light curves, the measured time delays were found to get highly biased with respect to the true time delays for high values of $s$, such as $s = 100\delta$. Hence in this work, we set $s = 10\delta$ by default. If the maximum absolute ratio noted above is $\geq$2, we consider smaller values of $s = 5\delta$ and $s = 2.5\delta$, until the maximum absolute ratio is <2 (with $s = 2.5\delta$ being the minimum value chosen even if the corresponding value of maximum absolute ratio is $\geq$2). However, it is possible that the time delay estimator can exhibit unstable behavior for low values of $s$, especially for relatively poor quality light curves. In such cases, we keep doubling the value of $s$ until we are able to reliably estimate the time delay and its uncertainty. For this purpose, it is useful to examine the $\overline{\chi}^2$ plot, to be described in Section~\ref{sec:normalized-chi-squared-plot}, over the range of trial time delays being considered for different choices of the value of $s$. We note that choosing from discrete values of $s$ in this manner and not optimizing its value as in \citet{Kumar2015} also lead to considerable saving of computational time.   

\subsection{Estimation of uncertainty}
\label{sec:time-delay-uncertainty}

As important as measuring the time delay itself is reliably estimating the uncertainty of the measured time delay. We introduce a `simple' uncertainty, which is inferred by applying the time delay measurement algorithm on synthetic light curves simulating only the measured time delay. Estimating `simple' uncertainty is relatively fast and we use this uncertainty estimate when testing the difference-smoothing algorithm on a large sample of TDC1 simulated light curves in Section~\ref{sec:crash-testing}. `Comprehensive' uncertainty, which is a refinement over the uncertainty of measured time delay as was estimated in \citet{Kumar2015}, is inferred by applying the algorithm on synthetic light curves simulating not only the measured time delay but also other time delays spaced uniformly at discrete values and spanning a sufficiently broad range around the measured time delay. In this work, we estimate `comprehensive' uncertainty in Section~\ref{sec:crash-testing} only for those TDC1 light curves for which the measured time delays are discrepant with the true time delays at >2$\sigma_i$ level when estimating `simple' uncertainty in order to test its robustness.  

\subsubsection{`Simple' uncertainty}
\label{sec:simple-uncertainty} 

We first apply the technique, using the same values of free parameters -- $\delta$ and $s$ -- employed for measuring the time delay from the observed light curves, on a large number $N_s$ of synthetic light curves having a simulated time delay equal to the measured time delay $\Delta t$. We compute the mean and standard deviation of the $N_s$ resulting values of time delays. The standard deviation gives us an estimate of the random error, whereas the departure of the mean from the simulated time delay, that was used in generating the synthetic light curves, gives us an estimate of the systematic error. By adding the random error and the systematic error in quadrature, we obtain a first estimate of the total error that we denote as ${\Delta\tau}_0$. In this work, we refer to this as `simple' uncertainty. We note that in applying the time delay estimator on $N_s$ synthetic light curves, the optimizer we employ might undergo catastrophic failure in a few cases. Hence, to avoid the overestimation of uncertainty we perform iterative 4$\sigma$ rejection of outliers among the time delay values measured from the synthetic light curves prior to the calculation of the random error and the systematic error. In this work, we have used $N_s$ = 500. For this choice of the number of synthetic light curves, the random error gets estimated to a precision of $1/\sqrt{2(N_s-1)} \sim$ 3 per cent and the systematic error gets estimated to a precision equalling $1/\sqrt{N_s} \sim$ 4 per cent of the magnitude of the random error \citep[e.g.][]{Taylor1997}. 

\subsubsection{`Comprehensive' uncertainty}
\label{sec:comprehensive-uncertainty} 

In order to adequately penalize for the `lethargy' of the time delay estimator \citep{Tewes2013a,Tewes2013,Eulaers2013,Kumar2013,Bonvin2017}, we also apply the technique on $N_s$ synthetic light curves for each value of simulated time delay differing from the measured time delay $\Delta t$ by $\pm\frac{m}{2}$ and $\pm m$, in each step updating the total error by adding the maximum obtained value of the random error and the maximum obtained absolute value of the systematic error in quadrature. Here $m$, as introduced previously, is the mean sampling of the light curves computed after excluding the seasonal gaps. In this work, we propose that the half-width of the range of simulated time delays, over which synthetic light curves need to be generated for the purpose of estimation of uncertainty of the measured time delay, should at least equal $m$. In general, we further extend this range in multiples of $\frac{m}{2}$ until the range of simulated time delays has a half-width of $n\left(\frac{m}{2}\right) \geq 2{\Delta\tau}_n$. This condition ensures that we have applied the time delay estimator on synthetic light curves having simulated time delays over a range which is at least as wide as the 95.4 per cent confidence interval implied by the final estimate of the total error ${\Delta\tau}_n$, which we refer to as `comprehensive' uncertainty. 

We note the differences in the approach followed to estimate `comprehensive' uncertainty with respect to that followed in \citet{Kumar2015}. In this work, the values of simulated time delays are uniformly spaced from the measured time delay in intervals of $m/2$, whereas in previous work each interval between the simulated time delays was chosen according to the recently updated estimate of the total error. Also in \citet{Kumar2015}, by not requiring the half-width of the range of time delays simulated by the synthetic light curves to at least equal $m$, the time-scale in which the variability features of the background quasar is resolvable in the observed light curves, the procedure was prone to the risk of underestimating the value of `comprehensive' uncertainty. We note that decreasing the interval between the sampled values of simulated time delay to smaller than $m/2$ is not found to significantly alter the estimate of `comprehensive' uncertainty. 

\subsection{Assessing the reliabilty of the measured time delay}
\label{sec:normalized-chi-squared-plot} 

To judge the reliability of the measured time delay and also to identify instances of catastrophic failure, we visually examine the merged light curve $M_i$ to see how well the variability features match between the two light curves. In addition to this, we find it useful to plot the values of $\overline{\chi}^2$ over the entire range of trial time delay values under consideration. In the case of high quality light curves, the minimum corresponding to the true time delay can be unambiguously identified. However, if we find multiple minima in the $\overline{\chi}^2$ plot whose characteristics are comparable to one another, as can happen in the case of marginal quality light curves, we flag the time delay measurement as being unreliable. 

\subsection{Correcting the measured time delay for systematic bias}
\label{sec:systematic-bias-correction} 

By applying the difference-smoothing algorithm on a large sample of TDC1 simulated light curves with known true time delays, as discussed in Section~\ref{sec:crash-testing}, we found that the difference-smoothing algorithm has an inherent tendency to measure the magnitude of time delay to be larger than that of the true time delay. Hence, we propose that the measured time delay be corrected according to the magnitude and sign of the systematic error obtained by applying the time delay estimator on synthetic light curves simulating the measured time delay, as discussed in Section~\ref{sec:simple-uncertainty}. This method of applying a correction to the measured time delay according to the systematic error, obtained during the uncertainty estimation procedure, is found to effectively eliminate the intrinsic systematic bias of the difference-smoothing algorithm, as demonstrated in Section~\ref{sec:crash-testing}.  

\section{Testing on TDC1 simulated light curves}
\label{sec:crash-testing}

The publicly available simulated light curves of TDC1 with known true time delays are arranged in five `rungs' having different sampling properties \citep[see][Table 1]{Liao2015}, in increasing order of difficulty. Whereas COSMOGRAIL\footnote{\url{http://cosmograil.epfl.ch/}}-like rung0 light curves have observing seasons of 8 month duration, light curves of all other LSST-like `rungs' have observing seasons of 4 month duration. Except rung4 light curves that have a cadence of 6 d, light curves of all other `rungs' have a cadence of 3 d. Whereas rung1 and rung4 light curves consist of ten observing seasons, light curves of the remaining `rungs' consist of five observing seasons. Except rung2 light curves which are uniformly sampled in time, light curves of all other `rungs' are unevenly sampled with a dispersion of 1 d. To quantify the performance of the difference-smoothing algorithm following the refinements proposed in the present work, we made use of more than 500 pairs of simulated light curves from each of the two most difficult `rungs' -- rung3 and rung4 -- of TDC1 selected according to their variability properties.

\subsection{Selection of light curves for analysis}
\label{sec:light-curves-selection} 

In general, the ease with which the time delay can be measured from a given pair of light curves and the precision of the measurement depend on the presence of a significant amount of variability with respect to the level of noise in both the light curves. Hence, for each observing season in the $A$ light curve we estimate noise by taking the standard deviation of $r_A(t_i)$ values for the epochs within the season and variability by taking the standard deviation of $A(t_i)$ values for the epochs within the season. Similarly, for each observing season in the $B$ light curve, we can obtain estimates of variability and noise. We can thus estimate for each observing season of $A$ and $B$ light curves a quantity, which we shall refer to as normalized seasonal variability, by dividing the estimate of variability by the estimate of noise. In this work, we analyse only those pairs of rung3 and rung4 TDC1 simulated light curves, in which at least one observing season of $A$ light curve and one observing season of $B$ light curve have normalized seasonal variability $\geq$2. 

\subsection{Measurement of time delays from TDC1 simulated light curves}
\label{sec:application-light-curves} 

As a first step, we visually inspect the light curves and mask epochs having extreme outliers from further analysis. We also try to assess if the light curves have extrinsic variations present, i.e. brightness variations that are uncorrelated between the two light curves, which as noted previously could arise due to microlensing by stars in the lensing galaxy. If in a certain observing season, one of the light curves exhibits large magnitude variation and the other light curve shows only little variation, this could be due to the particular observing season being affected by the presence of very fast extrinsic variations (assuming that the time delay between the light curves is no more than the length of the observing season, which is incidentally the case with all TDC1 light curves). Although difference-smoothing algorithm can adequately handle the presence of slow to moderately fast extrinsic variations in the light curves, the presence of observing seasons with very fast extrinsic variations can complicate the measurement of time delay. Hence, we completely mask any such observing season which hint the presence of very fast extrinsic variations from further analysis. 

In the absence of significant amount of extrinsic variations, we search for the time delay using the optimizer between $-$120 d and $+$120 d (120 d being the length of individual observing seasons for rung3 and rung4 light curves), setting the value of smoothing time-scale free parameter to $s = 100\delta$. We use the plot of $\overline{\chi}^2$ computed over the range of trial time delays to check if the minimum corresponding to the measured time delay can be unambiguously identified and also to find out if the optimizer had undergone catastrophic failure by getting trapped in a different minimum or a saddle point. Once the minimum corresponding to the time delay can be unambiguously identified, we limit the range of trial time delay values to be around the measured time delay based on visual inspection of $\overline{\chi}^2$ plot. In this instance of there not being significant amount of extrinsic variations, the maximum absolute ratio, i.e. the maximum of $\left|\frac{f_{A_i}}{\hat{\sigma}_{A_i}}\right|$ and $\left|\frac{f_{B_i}}{\hat{\sigma}_{B_i}}\right|$ (see Section~\ref{sec:synthetic-lightcurves}), will be $\lesssim$2 for a smoothing time-scale free parameter value of $s = 100\delta$. We then proceed to choose the value of $s$, as discussed in Section~\ref{sec:free-parameters}.

In the presence of significant amount of extrinsic variations, as before we search for the time delay using the optimizer between $-$120 d and $+$120 d, setting the value of smoothing time-scale free parameter to $s = 5\delta$, thus allowing for a sufficiently flexible model for extrinsic variations. Here again, if the minimum corresponding to the time delay can be unambiguously identified in the $\overline{\chi}^2$ plot, we proceed further restricting the range of trial time delay values to be around the measured time delay and then choosing the value of $s$, as discussed in Section~\ref{sec:free-parameters}. After measuring the time delay with the chosen value of free parameter $s$, we then estimate `simple' uncertainty and correct the measured time delay for systematic bias as discussed in Section~\ref{sec:simple-uncertainty} and Section~\ref{sec:systematic-bias-correction}, respectively. For those light curves, for which the measured time delays are discrepant with the true time delays at >2$\sigma_i$ level when estimating `simple' uncertainty, we estimate `comprehensive' uncertainty, as discussed in Section~\ref{sec:comprehensive-uncertainty}, in order to see to what extent the tension between the measured time delay and the true time delay gets alleviated in each case.

\begin{figure*}
	\includegraphics[width=\textwidth]{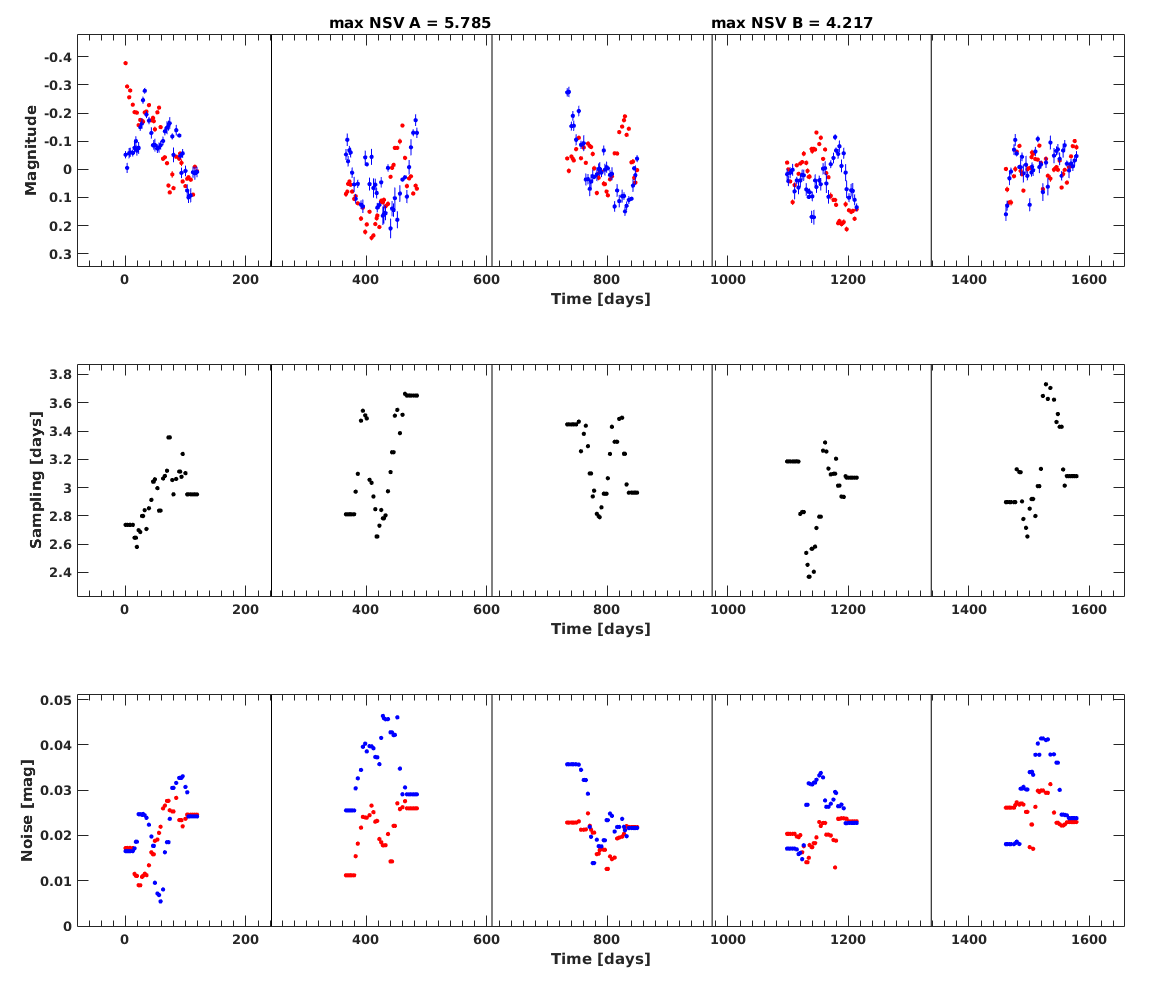}
    \caption{The top panel displays the $A$ and $B$ light curves corresponding to the filename `tdc1\_rung3\_double\_pair435.txt' in red and blue, respectively. The median magnitudes have been subtracted from both the light curves to enable easier visual comparison. The vertical lines separate the different observing seasons. The maximum values among the values of normalized seasonal variability calculated for all observing seasons for the $A$ and $B$ light curves are shown above the plot. The middle panel displays local estimates of mean sampling $m(t_i)$. The bottom panel displays the empirical estimates of noise in $A$ and $B$ light curves -- $\hat{\sigma}_{A_i}$ and $\hat{\sigma}_{B_i}$ -- in red and blue, respectively.}
    \label{fig:light-curves}
\end{figure*}

\begin{figure*}
	\includegraphics[width=\textwidth]{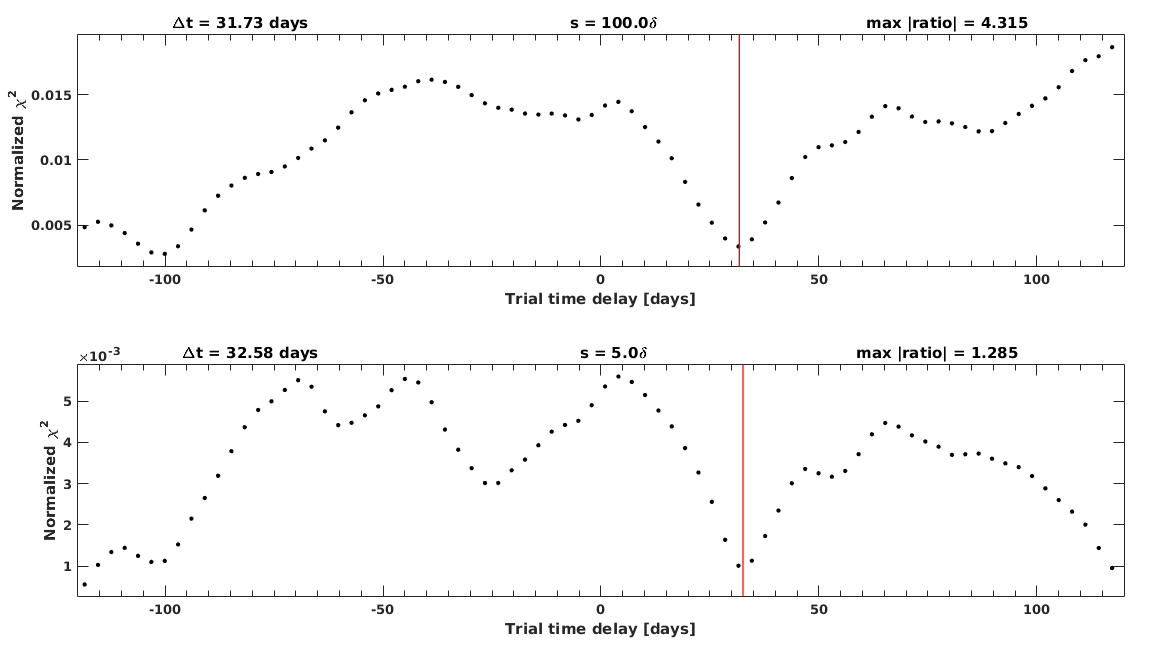}
    \caption{The bottom panel displays the plot of $\overline{\chi}^2$ computed for trial time delay values spaced at decorrelation length $\delta$ with the value of smoothing time-scale free parameter $s = 5\delta$, which unambiguously reveals the time delay to be $\sim$33 d with $A$ light curve leading $B$ light curve. The vertical red line indicates the time delay measured by the optimizer. The maximum absolute ratio (see Section~\ref{sec:free-parameters}) corresponding to the measured time delay is shown above the plot. The top panel shows the $\overline{\chi}^2$ plot corresponding to $s = 100\delta$.}
    \label{fig:normalized-chi-squared}
\end{figure*}

\begin{figure*}
	\includegraphics[width=\textwidth]{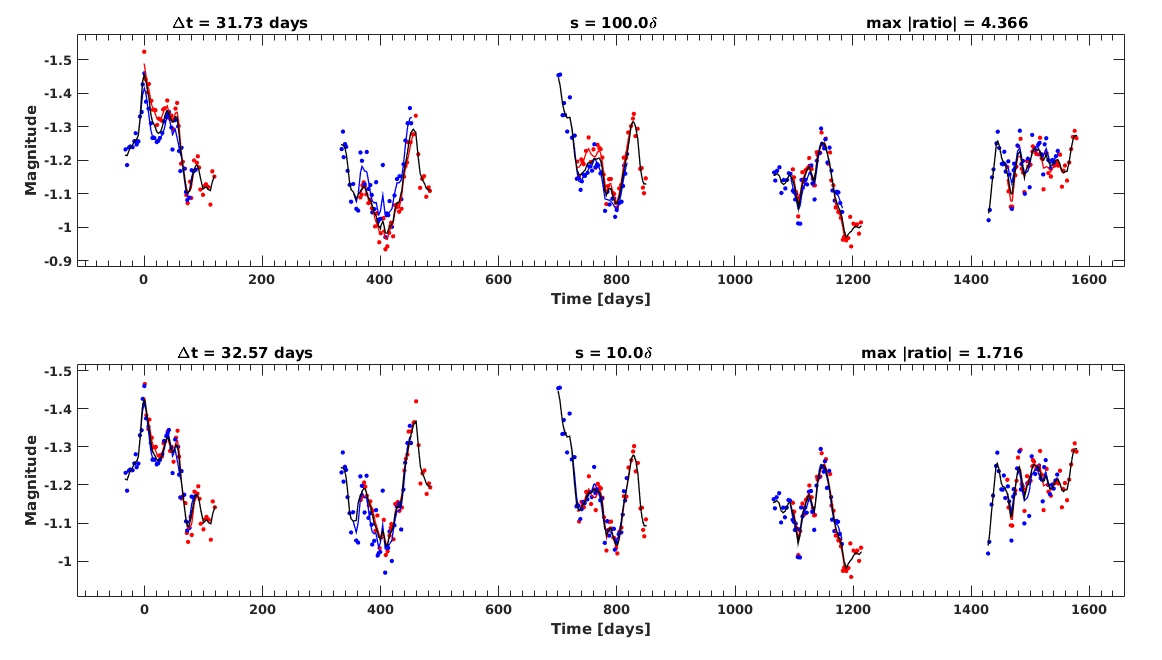}
    \caption{The bottom panel displays the plot of merged light curves $M_i$, when $A$ light curve is used as reference, for the time delay measured with $s = 10\delta$. Black lines connect $M(t)$ sampled at the epochs of $M_i$. Red and blue lines connect $M_A(t_i)$ and $M_B(t_i-\Delta t)$, respectively. The maximum absolute ratio (see Section~\ref{sec:free-parameters}) is shown above the plot. The top panel shows the corresponding plot for the time delay measured with $s = 100\delta$, from which the presence of significant amount of residual extrinsic variations can be clearly seen.}
    \label{fig:merged-light-curves}
\end{figure*}

\begin{figure}
	\includegraphics[width=\columnwidth]{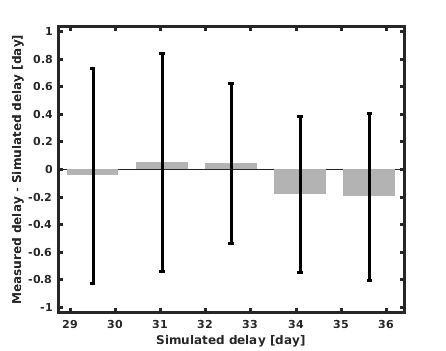}
    \caption{The systematic errors committed by the time delay estimator with $s = 10\delta$ on synthetic light curves having simulated time delays around the measured time delay (32.57 d) spaced at $m/2$ over a range having half-width of $m$ (3.06 d) are plotted as grey bars. The error bars denote the magnitude of the random errors. The `comprehensive' uncertainty (0.82 d) is computed by adding the maximum obtained value of a random error (0.79 d) and the maximum obtained absolute value of systematic error (0.20 d) in quadrature.}
    \label{fig:comprehensive-uncertainty}
\end{figure}

We illustrate the above process for one pair of TDC1 simulated light curves -- having filename `tdc1\_rung3\_double\_pair435.txt' -- which is displayed in Fig.~\ref{fig:light-curves}, along with estimates of local mean sampling $m(t_i)$ and empirical estimates of noise -- $\hat{\sigma}_{A_i}$ and $\hat{\sigma}_{B_i}$ -- that are used in generating synthetic light curves. 
Applying the time delay estimator with the value of $s = 100\delta$ reveals the presence of significant amount of differential extrinsic variations between the two light curves (top panels of Fig.~\ref{fig:normalized-chi-squared} and Fig.~\ref{fig:merged-light-curves}). Hence allowing for a flexible model of extrinsic variations, we search for the time delay with the value of $s = 5\delta$. The resulting $\overline{\chi}^2$ plot is shown in the bottom panel of Fig.~\ref{fig:normalized-chi-squared}, which unambiguously reveals the time delay to be $\sim$33 d with $A$ light curve leading $B$ light curve.
We fixed $s = 10\delta$, for which the maximum absolute ratio (see Section~\ref{sec:free-parameters}) was found to be 1.721 and measured the time delay to be 32.57 d. A plot of the merged light curves corresponding to $s = 10\delta$, when $A$ light curve is used as reference, is shown in the bottom panel of Fig.~\ref{fig:merged-light-curves}. Estimating `simple' uncertainty with the range of trial time delay values restricted to between 0 d and 70 d (based on visual inspection of the $\overline{\chi}^2$ plot) and correcting the measured time delay for systematic bias, the time delay between the two light curves is 32.58 $\pm$ 0.58 d, which is significantly discrepant with the true time delay of 31.18 d at the level of 2.41$\sigma_i$. Estimating `comprehensive' uncertainty (see Fig.~\ref{fig:comprehensive-uncertainty}) and correcting the measured time delay for systematic bias, the time delay between the two light curves is 32.52 $\pm$ 0.82 d (the value of time delay is slightly different from earlier observations due to the correction applied based on the estimate of the systematic error having an uncertainty on account of generating only a finite number of synthetic light curves, as noted in Section~\ref{sec:simple-uncertainty}), which is found to be in agreement with the true time delay (31.18 d) to well within 2$\sigma_i$.

\subsection{Calculation of TDC performance metrics}
\label{sec:tdc-performance-metrics}

The TDC performance metrics as defined in \citet{Dobler2015} and \citet{Liao2015} are summarized below for the convenience of the reader. The success fraction or efficiency $f$ of the time delay estimator is the fraction of light curves for which time delays have been submitted $N_{\mathrm{sub}}$ with respect to the total number of light curves $N$ available for analysis,
\begin{equation}
    f=\frac{N_{\mathrm{sub}}}{N}.
	\label{eq:success-fraction}
\end{equation} 
The goodness of fit between the measured time delays and the true time delays is quantified by standard reduced $\chi^2$ as
\begin{equation}
    \chi^2=\frac{1}{fN}{\sum_i\left(\frac{\tilde{\Delta t_i}-\Delta t_i}{\delta_i}\right)}^2,
	\label{eq:goodness-of-fit}
\end{equation}
where $\Delta t_i$ (defined to be positive) denote the true time delays, $\tilde{\Delta t_i}$ denote the measured time delays and $\delta_i$ denote the uncertainties of the measured time delays. The claimed precision $P$ of the time delay estimator is the average relative uncertainty per lens,
\begin{equation}
    P=\frac{1}{fN}\sum_i\frac{\delta_i}{\Delta t_i}.
	\label{eq:precision}
\end{equation} 
The accuracy or bias $A$ of the time delay estimator is the average fractional residual per lens,
\begin{equation}
    A=\frac{1}{fN}\sum_i\frac{\tilde{\Delta t_i}-\Delta t_i}{\Delta t_i}.
	\label{eq:accuracy}
\end{equation} 
The analogous metrics for each individual measurement are defined as
\begin{equation}
    \chi^2_i={\left(\frac{\tilde{\Delta t_i}-\Delta t_i}{\delta_i}\right)}^2,
	\label{eq:goodness-of-fit-individual}
\end{equation} 
\begin{equation}
    P_i=\frac{\delta_i}{\Delta t_i}
	\label{eq:precision-individual}
\end{equation} 
and
\begin{equation}
    A_i=\frac{\tilde{\Delta t_i}-\Delta t_i}{\Delta t_i}.
	\label{eq:accuracy-individual}
\end{equation} 
The uncertainties of $\chi^2$,$P$ and $A$ are calculated by taking the standard deviations of $\chi^2_i$,$P_i$ and $A_i$ values, respectively, and dividing by $\sqrt{fN}$.

\subsection{Results}
\label{sec:results}

\begin{figure*}
	\includegraphics[width=\textwidth]{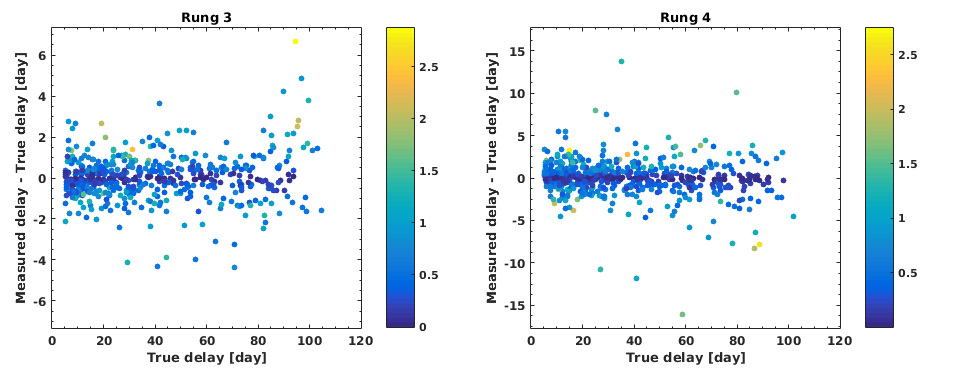}
    \caption{The differences between the measured time delays $\tilde{\Delta t_i}$ and the true time delays $\Delta t_i$ are plotted as a function of true time delay. The individual measurements are colour coded according to the values of $\left|\frac{\tilde{\Delta t_i}-\Delta t_i}{\delta_i}\right|$, where $\delta_i$ are the estimates of `simple' uncertainty.}
    \label{fig:truth-comparison}
\end{figure*}

Each TDC1 `rung' consists of 1024 light curves. A total number of 1264 light curves -- 594 from rung3 and 670 from rung4 -- satisfied the criterion for selection of light curves for analysis described in Section~\ref{sec:light-curves-selection}. Of those, we were able to successfully measure the time delays and estimate their respective `simple' uncertainties for a total of 1076 TDC1 light curves -- 517 from rung3 and 559 from rung4. The differences between the measured time delays $\tilde{\Delta t_i}$ and the true time delays $\Delta t_i$ are plotted as a function of true time delay in Fig.~\ref{fig:truth-comparison}, where the individual measurements are colour coded according to the values of $\left|\frac{\tilde{\Delta t_i}-\Delta t_i}{\delta_i}\right|$, where $\delta_i$ are the estimates of `simple' uncertainty. 
We see that all the measured time delays agree with the true time delays to within 3$\sigma_i$. The TDC performance metrics calculated with all the measurements and after selecting only those measurements that have empirical precision $\left|\frac{\delta_i}{\tilde{\Delta t_i}}\right|$ of $\leq$20 per cent and $\leq$10 per cent are presented in Table~\ref{tab:tdc-metrics-all}.
\begin{table*}
	\centering
	\caption{TDC performance metrics for the difference-smoothing algorithm calculated with all the measured time delays and after selecting only those measurements having empirical precision of $\leq$20 per cent and $\leq$10 per cent.}
	\label{tab:tdc-metrics-all}
	\begin{tabular}{lccccc} 
		\hline
		Selection & Rung & f & $\chi^2$ & P & A\\
		\hline
		All measurements & 3 & 0.51 & 0.460 $\pm$ 0.032 & 0.064 $\pm$ 0.004 & 0.000 $\pm$ 0.002\\
		All measurements & 4 & 0.55 & 0.399 $\pm$ 0.030 & 0.119 $\pm$ 0.005 & 0.008 $\pm$ 0.004\\
		Precision $\leq$20 per cent & 3 & 0.48 & 0.470 $\pm$ 0.034 & 0.046 $\pm$ 0.002 & 0.001 $\pm$ 0.002\\
		Precision $\leq$20 per cent & 4 & 0.47 & 0.399 $\pm$ 0.034 & 0.079 $\pm$ 0.002 & 0.005 $\pm$ 0.003\\
		Precision $\leq$10 per cent & 3 & 0.43 & 0.480 $\pm$ 0.037 & 0.037 $\pm$ 0.001 & 0.002 $\pm$ 0.001\\
		Precision $\leq$10 per cent & 4 & 0.34 & 0.415 $\pm$ 0.043 & 0.055 $\pm$ 0.001 & 0.001 $\pm$ 0.002\\
		\hline
	\end{tabular}
\end{table*} 
We see that the measurements achieve sub-percent accuracy and do not incur significant bias. We note that failing to correct each of the measured time delays according to the magnitude and sign of the systematic error, as described in Section~\ref{sec:systematic-bias-correction}, leads to significant bias of $\sim$1 per cent and $\sim$2 per cent for rung3 and rung4 light curves, respectively.

The TDC1 simulated light curves have true time delays between 5 d and 120 d, whereas the true time delays of the rung3 and rung4 light curves for which we are able to successfully measure the time delays and estimate their respective `simple' uncertainties range between 5 d and 105 d. We divide this range into five bins, each spanning 20 d, to investigate the possibility of systematic bias being dependent on the magnitude of true time delay. The TDC performance metrics for each of these true time delay bins are presented in Table~\ref{tab:tdc-metrics-binned}, using only those measurements having empirical precision of $\leq$10 per cent. We note that the value of $f$ for each true time delay bin has been calculated with respect to the total number of light curves that are available within that true time delay range. 
\begin{table*}
	\centering
	\caption{TDC performance metrics for the difference-smoothing algorithm calculated for five different true time delay bins, selecting only those measurements that have empirical precision of $\leq$10 per cent.}
	\label{tab:tdc-metrics-binned}
	\begin{tabular}{lccccc}
		\hline
		True time delay range & Rung & f & $\chi^2$ & P & A\\
		\hline
		5 d $\leq \Delta t$ < 25 d & 3 & 0.34 & 0.482 $\pm$ 0.062 & 0.054 $\pm$ 0.002 & 0.002 $\pm$ 0.004\\
		5 d $\leq \Delta t$ < 25 d & 4 & 0.17 & 0.579 $\pm$ 0.131 & 0.076 $\pm$ 0.002 & 0.003 $\pm$ 0.007\\
		25 d $\leq \Delta t$ < 45 d & 3 & 0.50 & 0.495 $\pm$ 0.070 & 0.036 $\pm$ 0.002 & $-$0.000 $\pm$ 0.002\\
		25 d $\leq \Delta t$ < 45 d & 4 & 0.46 & 0.353 $\pm$ 0.060 & 0.056 $\pm$ 0.002 & 0.004 $\pm$ 0.003\\
		45 d $\leq \Delta t$ < 65 d & 3 & 0.53 & 0.302 $\pm$ 0.039 & 0.027 $\pm$ 0.002 & 0.003 $\pm$ 0.002\\
		45 d $\leq \Delta t$ < 65 d & 4 & 0.51 & 0.357 $\pm$ 0.058 & 0.046 $\pm$ 0.002 & 0.003 $\pm$ 0.003\\
		65 d $\leq \Delta t$ < 85 d & 3 & 0.58 & 0.393 $\pm$ 0.054 & 0.023 $\pm$ 0.002 & $-$0.002 $\pm$ 0.002\\
		65 d $\leq \Delta t$ < 85 d & 4 & 0.54 & 0.332 $\pm$ 0.075 & 0.047 $\pm$ 0.002 & $-$0.002 $\pm$ 0.004\\
		85 d $\leq \Delta t$ < 105 d & 3 & 0.43 & 0.925 $\pm$ 0.244 & 0.022 $\pm$ 0.002 & 0.009 $\pm$ 0.003\\
		85 d $\leq \Delta t$ < 105 d & 4 & 0.32 & 0.595 $\pm$ 0.247 & 0.048 $\pm$ 0.003 & $-$0.009 $\pm$ 0.006\\
		\hline
	\end{tabular}
\end{table*}
We find that the measurements are significantly biased at the level of 0.9 $\pm$ 0.3 per cent for the true time delay bin ranging between 85 d and 105 d for rung3 light curves. This is presumably because we had to use high values of the free parameter $s$, corresponding to rigid models of differential extrinsic variations, for many of the light curves in order to be able to successfully measure their time delays and estimate the respective `simple' uncertainties, on account of the narrow overlap (15--35 d) in each observing season between light curves $A$ and $B$ for the high time delay values of that bin. As discussed previously in Section~\ref{sec:free-parameters}, the measured time delays can get highly biased with respect to the true time delays for high values of $s$. Hence, we further select only those measurements for which we had used the values of $s \leq 10\delta$ and present the TDC performance metrics for the different true time delay bins in Table~\ref{tab:tdc-metrics-binned-s-leq-10-delta}.
\begin{table*}
	\centering
	\caption{TDC performance metrics for the difference-smoothing algorithm calculated for five different true time delay bins, selecting only those measurements having empirical precision of $\leq$10 per cent and which are carried out with the value of smoothing time-scale free parameter $s \leq$ 10$\delta$.}
	\label{tab:tdc-metrics-binned-s-leq-10-delta}
	\begin{tabular}{lccccc}
		\hline
		True time delay range & Rung & f & $\chi^2$ & P & A\\
		\hline
		5 d $\leq \Delta t$ < 25 d & 3 & 0.34 & 0.482 $\pm$ 0.062 & 0.054 $\pm$ 0.002 & 0.002 $\pm$ 0.004\\
		5 d $\leq \Delta t$ < 25 d & 4 & 0.17 & 0.579 $\pm$ 0.131 & 0.076 $\pm$ 0.002 & 0.003 $\pm$ 0.007\\
		25 d $\leq \Delta t$ < 45 d & 3 & 0.49 & 0.499 $\pm$ 0.071 & 0.035 $\pm$ 0.002 & $-$0.000 $\pm$ 0.002\\
		25 d $\leq \Delta t$ < 45 d & 4 & 0.46 & 0.353 $\pm$ 0.060 & 0.056 $\pm$ 0.002 & 0.004 $\pm$ 0.003\\
		45 d $\leq \Delta t$ < 65 d & 3 & 0.53 & 0.306 $\pm$ 0.039 & 0.027 $\pm$ 0.002 & 0.002 $\pm$ 0.002\\
		45 d $\leq \Delta t$ < 65 d & 4 & 0.49 & 0.354 $\pm$ 0.060 & 0.044 $\pm$ 0.002 & 0.002 $\pm$ 0.003\\
		65 d $\leq \Delta t$ < 85 d & 3 & 0.54 & 0.394 $\pm$ 0.057 & 0.021 $\pm$ 0.001 & $-$0.001 $\pm$ 0.002\\
		65 d $\leq \Delta t$ < 85 d & 4 & 0.35 & 0.353 $\pm$ 0.098 & 0.039 $\pm$ 0.002 & 0.002 $\pm$ 0.003\\
		85 d $\leq \Delta t$ < 105 d & 3 & 0.25 & 0.741 $\pm$ 0.252 & 0.018 $\pm$ 0.001 & 0.005 $\pm$ 0.003\\
		85 d $\leq \Delta t$ < 105 d & 4 & 0 & $-$ & $-$ & $-$\\
		\hline
	\end{tabular}
\end{table*}
We now find that the measurements are no longer significantly biased for any of the true time delay bins.

We now proceed to compare the TDC performance metrics obtained in this work with those of the best perfoming TDC1 submissions. For this purpose, we use only those light curves that have true time delays $\Delta t_i \geq 10$ d as was performed in \citet{Liao2015}, in addition to using only those measurements made with $s \leq 10 \delta$ and having empirical precision of $\leq$10 per cent. We have presented the resulting metrics in Table~\ref{tab:tdc-metrics-comparison-with-tdc1} along with the performance metrics of the TDC1 submissions for rung3 and rung4 that achieved sub-percent accuracy and catastrophic failure rate of $\leq$5 per cent, as listed in table 5 of \citet{Liao2015}. These metrics have been calculated after rejection of catastrophic outliers, which are defined as those measurements for which $\left|\tilde{\Delta t_i}-\Delta t_i\right|$ > 3.3$\delta_i$.  
\begin{table*}
	\centering
	\caption{TDC performance metrics for the difference-smoothing algorithm calculated using only light curves having true time delays $\Delta t_i \geq$ 10 d, for which the measurements were carried out with $s \leq 10\delta$ and have empirical precision of $\leq$10 per cent, compared with the metrics of the TDC1 submissions for rung3 and rung4 that achieved sub-percent accuracy and catastrophic failure rate of $\leq$5 per cent \citep[see][table 5]{Liao2015}. $X$ denotes the fraction of measurements that are not catastrophic outliers.}
	\label{tab:tdc-metrics-comparison-with-tdc1}
	\begin{tabular}{lcccccc}
		\hline
		Method & Rung & f$_{3.3\sigma}$ & $\chi^2_{3.3\sigma}$ & P$_{3.3\sigma}$ & A$_{3.3\sigma}$ & X\\
		\hline
		This work & 3 & 0.44 & 0.447 $\pm$ 0.034 & 0.035 $\pm$ 0.001 & 0.001 $\pm$ 0.001 & 1.0\\
		This work & 4 & 0.32 & 0.405 $\pm$ 0.043 & 0.055 $\pm$ 0.001 & 0.003 $\pm$ 0.002 & 1.0\\
		PyCS-sdi-vanilla-dou-full & 3 & 0.3 & 0.813 $\pm$ 0.074 & 0.068 $\pm$ 0.006 & $-$0.004 $\pm$ 0.006 & 1.0\\ 
		PyCS-sdi-vanilla-dou-full & 4 & 0.21 & 0.804 $\pm$ 0.096 & 0.098 $\pm$ 0.015 & 0.005 $\pm$ 0.006 & 0.99\\
		PyCS-spl-vanilla-dou-full & 3 & 0.3 & 0.494 $\pm$ 0.057 & 0.042 $\pm$ 0.003 & $-$0.001 $\pm$ 0.003 & 1.0\\ 
		PyCS-spl-vanilla-dou-full & 4 & 0.21 & 0.665 $\pm$ 0.065 & 0.045 $\pm$ 0.003 & 0.001 $\pm$ 0.003 & 1.0\\   
		Jackson-manchester2\_0\_3\_4 & 3 & 0.34 & 1.165 $\pm$ 0.099 & 0.036 $\pm$ 0.001 & 0.002 $\pm$ 0.003 & 0.98\\
		JPL & 3 & 0.28 & 1.28 $\pm$ 0.11 & 0.051 $\pm$ 0.004 & 0.007 $\pm$ 0.007 & 0.95\\
		Hojjati-Stark & 3 & 0.18 & 0.78 $\pm$ 0.12 & 0.06 $\pm$ 0.004 & $-$0.003 $\pm$ 0.005 & 0.96\\   
		Hojjati-Stark & 4 & 0.16 & 0.89 $\pm$ 0.14 & 0.07 $\pm$ 0.004 & 0.002 $\pm$ 0.005 & 0.98\\ 
		\hline
	\end{tabular}
\end{table*} 
We find that following the refinements proposed in this work, the TDC performance metrics for the difference-smoothing algorithm are competitive with those of the best performing TDC1 submissions. It is worth noting that the refined procedure is sufficiently robust to be able to avoid the presence of catastrophic outliers among the measurements, as was achieved by only the `PyCS' team during TDC1 \citep{Liao2015,Bonvin2016}

In order to test the robustness of `comprehensive' uncertainty, as revised in this work (Section~\ref{sec:comprehensive-uncertainty}), we carried out their estimates for those light curves (totaling five in rung3 and four in rung4) for which the measured time delays were in tension with the true time delays at >2$\sigma_i$ level when estimating `simple' uncertainty. The results for those light curves are presented in Table~\ref{tab:interesting_cases}.
\begin{table*}
	\centering
	\caption{Testing the robustness of `comprehensive' uncertainty using those light curves for which the measured time delays $\tilde{\Delta t}_i$ and the true time delays ${\Delta t}_i$ disagreed at >2$\sigma_i$ level when estimating `simple' uncertainty. Each of the time delays has a positive or a negative sign according to whether light curve $A$ leads light curve $B$ or vice versa.} 
	\label{tab:interesting_cases}
	\begin{tabular}{lccc}
		\hline
		Filename of light curves & $\Delta t_i$ & $\tilde{\Delta t}_i$ $\pm$ `simple' $\delta_i$ (Discrepancy) & $\tilde{\Delta t}_i$ $\pm$ `comprehensive' $\delta_i$ (Discrepancy)\\
                \hline
		tdc1\_rung3\_double\_pair143.txt & $-$94.59 d & $-$101.25 $\pm$ 2.32 d (2.87$\sigma_i$) & $-$101.32 $\pm$ 3.37 d (2.00$\sigma_i$)\\
		tdc1\_rung3\_double\_pair435.txt & 31.18 d & 32.58 $\pm$ 0.58 d (2.41$\sigma_i$) & 32.52 $\pm$ 0.82 d (1.63$\sigma_i$)\\
		tdc1\_rung3\_double\_pair658.txt & $-$95.15 d & $-$97.68 $\pm$ 1.19 d (2.13$\sigma_i$) & $-$97.77 $\pm$ 1.37 d (1.91$\sigma_i$)\\
		tdc1\_rung3\_quad\_pair28B.txt & 95.6 d & 98.43 $\pm$ 1.39 d (2.04$\sigma_i$) & 98.26 $\pm$ 1.58 d (1.68$\sigma_i$)\\
		tdc1\_rung3\_quad\_pair64B.txt & 19.17 d & 21.83 $\pm$ 1.30 d (2.05$\sigma_i$) & 21.77 $\pm$ 1.69 d (1.54$\sigma_i$)\\
		tdc1\_rung4\_double\_pair524.txt & 37.54 d & 40.33 $\pm$ 1.23 d (2.27$\sigma_i$) & 40.34 $\pm$ 1.42 d (1.97$\sigma_i$)\\
		tdc1\_rung4\_double\_pair540.txt & $-$88.81 d & $-$80.97 $\pm$ 3.02 d (2.60$\sigma_i$) & $-$80.74 $\pm$ 4.81 d (1.68$\sigma_i$)\\
		tdc1\_rung4\_quad\_pair3B.txt & 16.47 d & 12.69 $\pm$ 1.84 d (2.05$\sigma_i$) & 12.68 $\pm$ 2.69 d (1.41$\sigma_i$)\\
		tdc1\_rung4\_quad\_pair12B.txt & 14.98 d & 18.19 $\pm$ 1.17 d (2.74$\sigma_i$) & 18.11 $\pm$ 1.60 d (1.96$\sigma_i$)\\  
		\hline
	\end{tabular}
\end{table*} 
We find that with `comprehensive' uncertainty estimates, the discrepancy between the measured time delays and the true time delays is no more than $\sim$2$\sigma_i$ level for any of these light curves, illustrating the robustness of the refined procedure to estimate `comprehensive' uncertainty. We note that the small differences in the time delays between the last two columns of Table~\ref{tab:interesting_cases} is due to the correction applied to each measurement for removing systematic bias (as described in Section~\ref{sec:systematic-bias-correction}) having an uncertainty on account of generating only a finite number of synthetic light curves, as discussed in Section~\ref{sec:simple-uncertainty}. The MATLAB codes used in this work and the detailed results obtained by applying them on rung3 and rung4 simulated light curves of TDC1 are made publicly available through GitHub\footnote{\url{https://github.com/rathnakumars/difference-smoothing}}, in order to aid reproducibility efforts and for wider use by the community.           

\section{Conclusion}          
\label{sec:conclusion}

In this work, we have introduced refinements to the difference-smoothing algorithm for measurement of time delay from the light curves of the images of a gravitationally lensed quasar. The refinements mainly consist of a more pragmatic approach to choose the smoothing time-scale free parameter, generation of more realistic synthetic light curves for estimation of time delay uncertainty and the use of $\overline{\chi}^2$ plot to assess the reliability of a time delay measurement as well as to identify instances of catastrophic failure of the time delay estimator. Applying the difference-smoothing algorithm on a large sample of simulated light curves from the two most difficult `rungs' -- rung3 and rung4 -- of the first edition of Strong Lens Time Delay Challenge (TDC1) revealed the technique to have an inherent tendency to measure the magnitudes of time delays to be larger than the true values of time delays at the level of $\sim$1 per cent and $\sim$2 per cent for rung3 and rung4 light curves, respectively. However, this systematic bias was found to be eliminated by applying a correction to each measured time delay according to the magnitude and sign of the systematic error obtained by applying the time delay estimator on synthetic light curves simulating the measured time delay. As a result of the refinements proposed in this work, the TDC performance metrics of the difference-smoothing algorithm were found to be competitive with the corresponding metrics of the best performing TDC1 submissions for both the tested `rungs'. The refined procedure was also found to be sufficiently robust to avoid the presence of catastrophic outliers among the measurements, as had been achieved by only one team during TDC1. 

In testing the difference-smoothing algorithm on a large sample of simulated light curves from TDC1, we estimated `simple' uncertainty for each measured time delay, which is based on applying the time delay estimator on synthetic light curves simulating only the measured time delay. In this work, we also introduced refinements to the procedure for estimating `comprehensive' uncertainty, which is based on applying the time delay estimator on synthetic light curves simulating time delays in a sufficiently broad range around the measured time delay, with respect to the minimum range of simulated time delays and their values being uniformly spaced from one another. The robustness of `comprehensive' uncertainty was tested using those TDC1 light curves, for which the measured time delays were found to be in tension with the true time delays at >2$\sigma_i$ level when estimating `simple' uncertainty. We found that all the measured time delays agree with the true time delays to within $\sim$2$\sigma_i$ level when estimating `comprehensive' uncertainty and thus confirming the robustness of the refined procedure to estimate `comprehensive' uncertainty. The MATLAB codes used in this work along with the detailed results obtained have been made publicly available.       

\section*{Acknowledgements} 
We acknowledge useful discussions with Shashikiran Ganesh and Samuel Johnson. The organizers of Strong Lens Time Delay Challenge are thanked for providing the simulated data and the truth files. We acknowledge extensive use of the computational server of Astronomy \& Astrophysics Division and the Vikram-100 HPC server of Physical Research Laboratory. We thank the referee Nobuhiro Okabe for a constructive report, which helped to improve the presentation of this work. The financial support from Science and Engineering Research Board, Department of Science \& Technology, Government of India, through fellowship reference number PDF/2016/003848 is gratefully acknowledged.

%%%%%%%%%%%%%%%%%%%%%%%%%%%%%%%%%%%%%%%%%%%%%%%%%%



%\bibliographystyle{mnras}
%\bibliography{crash_testing_paper} % if your bibtex file is called example.bib


%%%%%%%%%%%%%%%%%%%%%%%%%%%%%%%%%%%%%%%%%%%%%%%%%%

% Don't change these lines
\bsp	% typesetting comment
\label{lastpage}
\end{document}